\documentclass[11pt]{elsarticle}
\usepackage{geometry}                
\usepackage{graphicx}
\usepackage{amssymb}
\usepackage{epstopdf}
\usepackage{subfigure}

\usepackage{anysize}

\begin{document}

\begin{frontmatter}

\title{On plant roots logical gates}

\author[1]{Andrew Adamatzky}
\address[1]{Unconventional Computing Centre, UWE, Bristol, UK}

\author[3,1]{Georgios Sirakoulis}
\address[3]{Department of Electrical \& Computer Engineering,  
Democritus University of Thrace, Xanthi, Greece}

\author[4,1]{Genaro J. Mart{\'i}nez}
\address[4]{Superior School of Computer Sciences, National Polytechnic Institute, Mexico}

\author[2]{Frantisek Balu\v{s}ka}
\address[2]{Institute of Cellular and Molecular Botany, University of Bonn, Germany}

\author[5]{Stefano Mancuso}
\address[5]{International Laboratory of Plant Neurobiology, University of Florence, Italy}



\begin{abstract}
Theoretical constructs of logical gates implemented with plant roots are morphological computing asynchronous devices. 
Values of Boolean variables are represented by plant roots. A presence of a plant root at a given site symbolises the 
logical {\sc True}, an absence the logical {\sc False}. Logical functions are calculated via interaction between roots. 
 Two types of two-inputs-two-outputs gates are proposed: a  gate $\langle x, y \rangle \rightarrow \langle xy, x+y \rangle$ 
 where root apexes are guided by gravity and a gate  $\langle x, y \rangle \rightarrow \langle \overline{x}y, x \rangle$ where 
 root apexes are guided by humidity.  We propose a design of binary half-adder 
 based on the  gates.
\end{abstract}


\end{frontmatter}

\section{Introduction}

\begin{figure}[!bp] 
\centering
\includegraphics[scale=0.4]{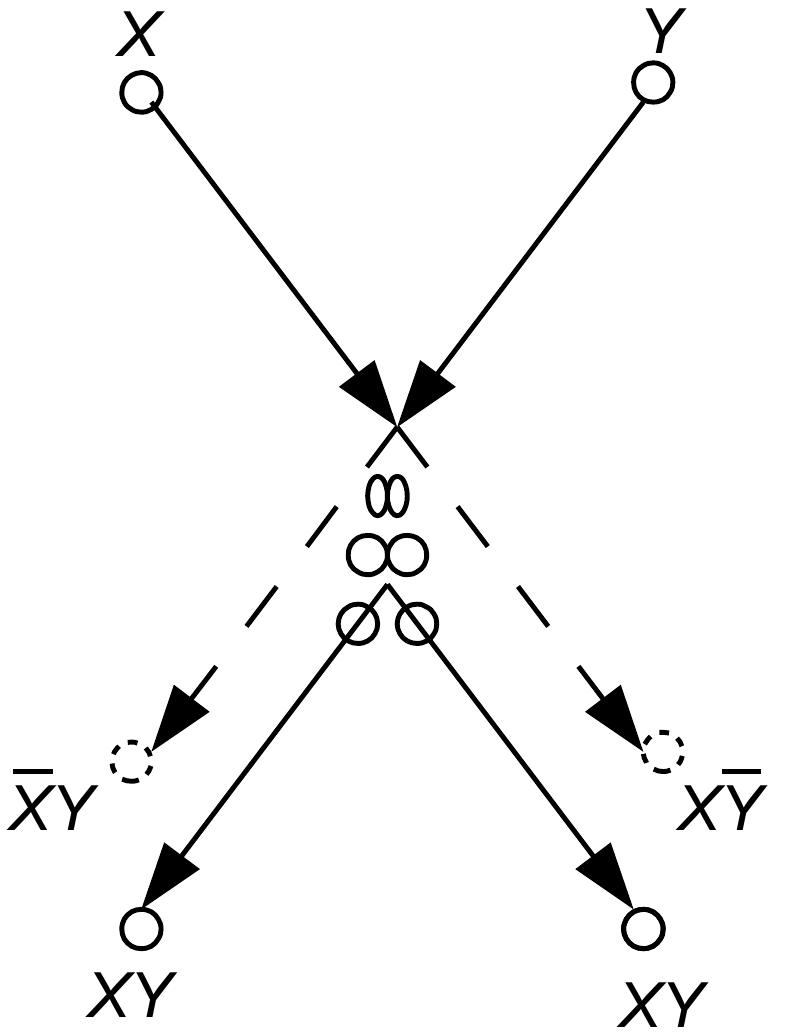}
\caption{Margolus gate: collision between soft balls.}
\label{margolus}
\end{figure}

A  collision-based computation, emerged from Fredkin-Toffoli conservative logic~\cite{fredkin2002conservative}, employs mobile compact finite patterns, which implement computation while interacting with each other~\cite{adamatzky2002collision}. Information values (e.g. truth values of logical variables) are given by either absence or presence of the localisations or other parameters of the localisations. The localisations travel in space and perform computation when they collide with each other.  Almost any part of the medium space can be used as a wire.  The localisations undergo transformations, they change velocities, form bound state,
{and} annihilate or fuse when they interact with other localisations. Information values of localisations are transformed as a result of collision and thus a computation is implemented.  

The concept of the collision-based logical gates is best illustrated using a gate based on collision between two soft balls, the Margolus gate~\cite{margolus2002universal}, shown in Fig.~\ref{margolus}. Logical value $x=1$ 
is given by a ball presented in input trajectory marked $x$, and $x=0$ by absence of the ball in the input trajectory $x$; the same applies to $y=1$ and $y=0$. When two balls approaching the collision gate along 
paths $x$ and $y$ collide, {they} compress but then spring back and reflect. The balls come out along the paths marked $xy$. If only one ball approaches the gate, for inputs $x=1$ and $y=0$ or $x=0$ and $y=1$, 
the balls exits the gate via path $x\overline{y}$ (for input $x=1$ and $y=0$) or $\overline{x}y$ (for input $x=0$ and $y=1$). 

{The} designed experimental prototypes of logical gates, circuits and binary adders {employ} 
interaction of wave-fragments in light-sensitive Belousov-Zhabotinsky media~\cite{costello2005experimental}, swarms of soldier crabs~\cite{gunji2011robust}, growing lamellipodia of slime mould {\emph Physarum polycephalum}~\cite{tsuda2004robust, adamatzky2010slime}, crystallisation patterns in `hot ice'~\cite{adamatzky2009hot}, peristaltic waves in protoplasmic tubes~\cite{adamatzky2014slime}, {and} jet streams in fluidic devices~\cite{morgan2016simple}, or as competing patterns propagation in channels of communication with a Life-like CA~\cite{martinez2010computation}. These prototypes suffer from various disadvantages. For example, wave-fragments in Belousov-Zhabotinsky medium are short-living and difficult to control (they are prone to expansion or contraction), slime mould 
protoplasmic tubes lack stability and exhibit tendency to uncontrolled branching, swarms of soldier crabs might behave chaotically and the gate requires a bulky setup. Another problem is a synchronisation. When Boolean values are represented by localised, finite size, 
patterns -- the accuracy of synchronisation depends on the size of the patterns. For example,  
if two  wave-fragments in Belousov-Zhabotinsky medium collide not `perfectly' but 
with an offset more than a half-wave length the output of the gate will be ineligible.  Thus we aimed to find physical or biological analogs where signals are well controlled and stable and large errors in synchronisation are allowed.

Plant roots could offer us a viable alternative. Plant roots could perform a computation by the following general mechanisms~\cite{baluvska2009root}: root tropisms stimulated via attracting and repelling spatially extended stimuli~\cite{yokawa2014binary}; 
morphological adaptation of root system architecture to attracting and repelling spatially extended stimuli~\cite{baluvska2004root}; wave-like propagation of information along the root bodies via plant-synapse networks~\cite{brenner2006plant}; 
patterns of which correlate with their environmental stimuli~\cite{mazzolai2008inspiration, mazzolai2010plant}; competition and entrainment of oscillations in their bodies~\cite{baluvska2010root}.  Computation results are represented by the topology of root apex trajectories which are preserved in a physical location of the root. The computing circuits proposed receive input signals on both inputs at the same time, synchronously, the signals are `desynchronised' en route due to different lengths of input channels.

\section{Gravity gates}

We propose gates made of channels. The roots grow inside the channels. The roots apexes navigate along the channel using mechanical, 
acoustic~\cite{gagliano2012towards} and visual~\cite{mo2015and, baluvska2016vision} means.
Root apexes exhibit a positive gravitropism~\cite{darwin1899geotropism, pfeffer1894geotropic,  baluvska1996gravitropism,baluvska1997root, masi2008electrical}. 
Thus being placed in a geometrically constraint environment  of a channel, a root grows along the same direction as the gravity force.

\begin{figure}[!tbp] 
\centering
\subfigure[]{\includegraphics[scale=0.3]{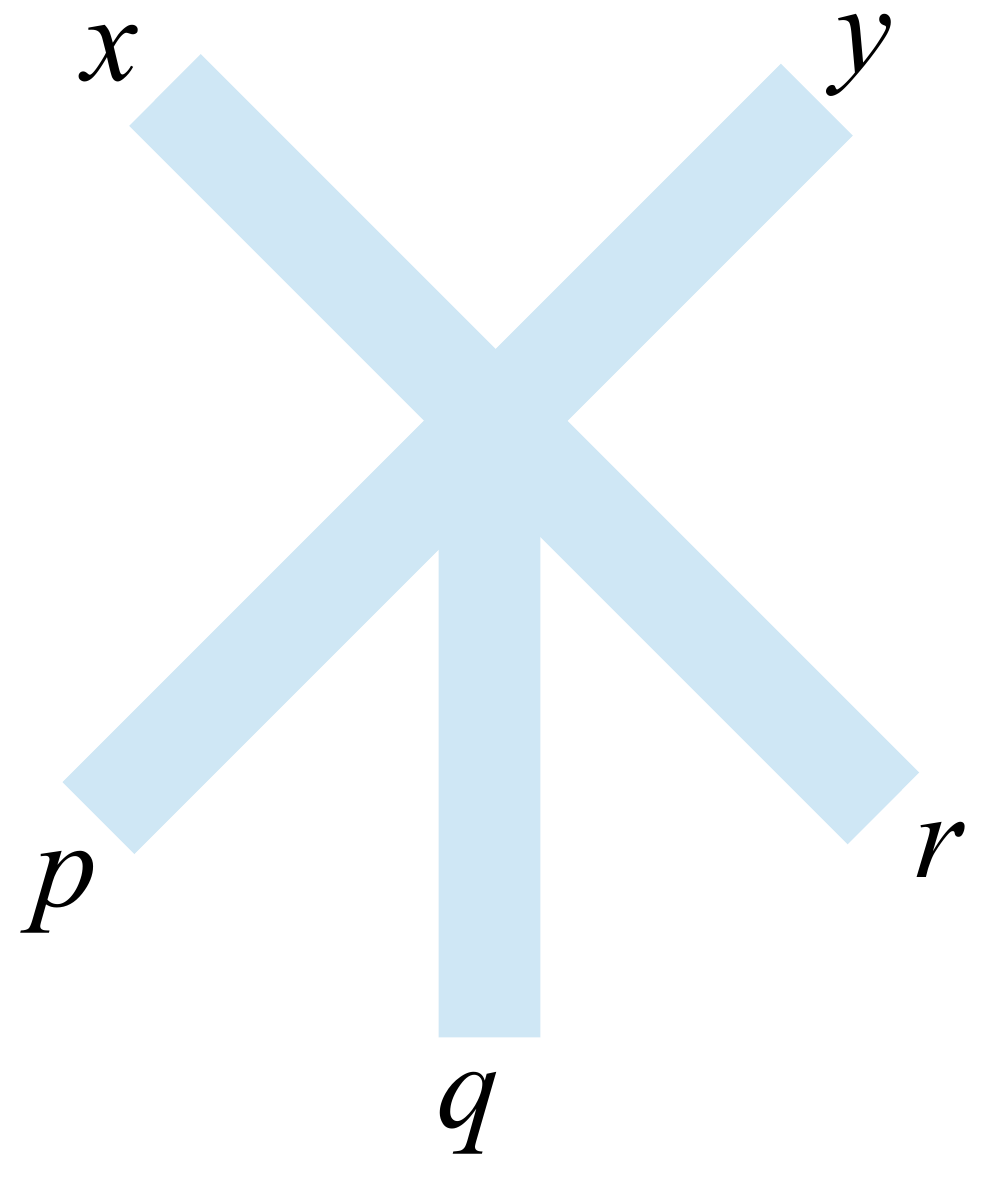}}
\subfigure[]{\includegraphics[scale=0.3]{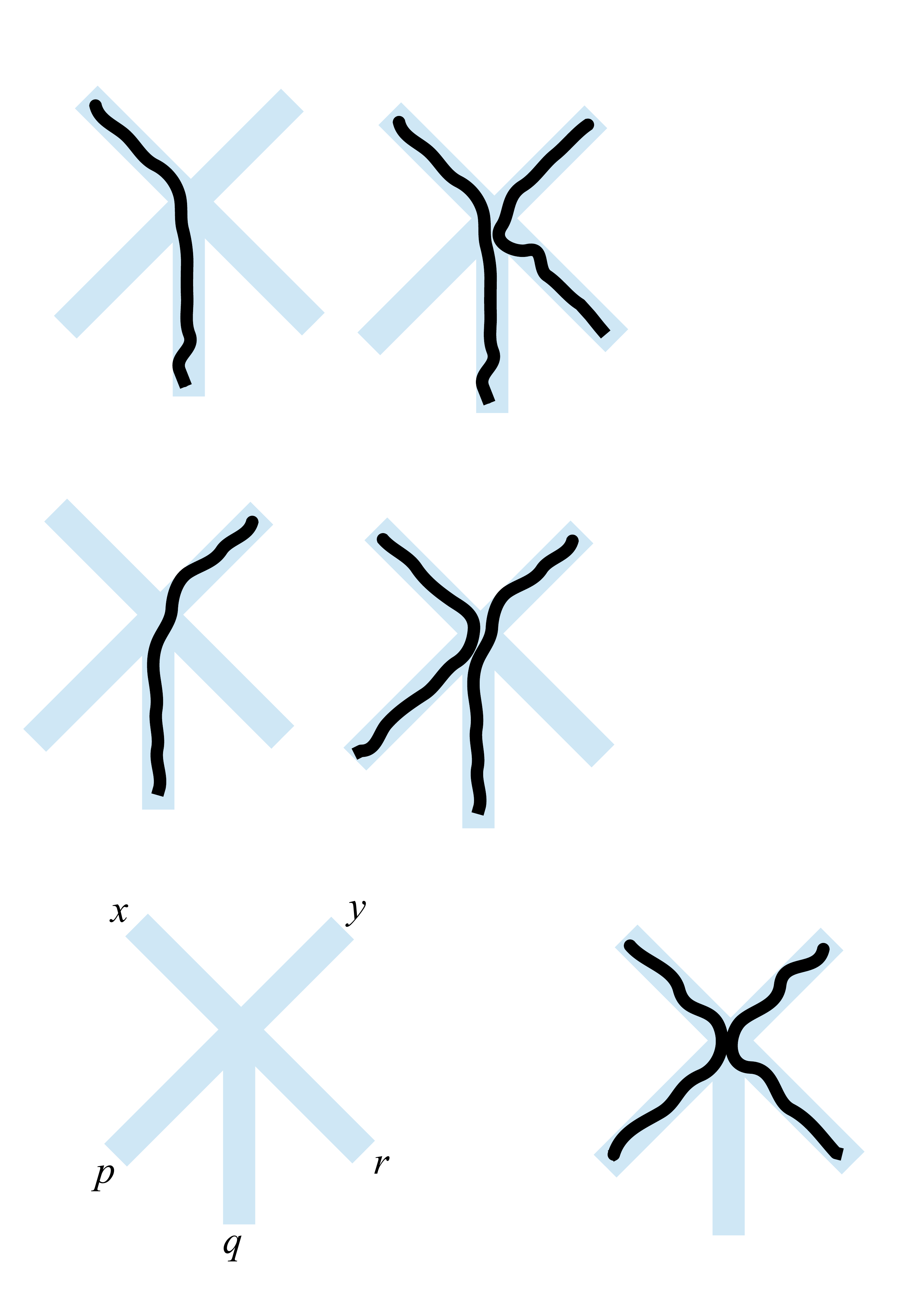}}
\subfigure[]{\includegraphics[scale=0.3]{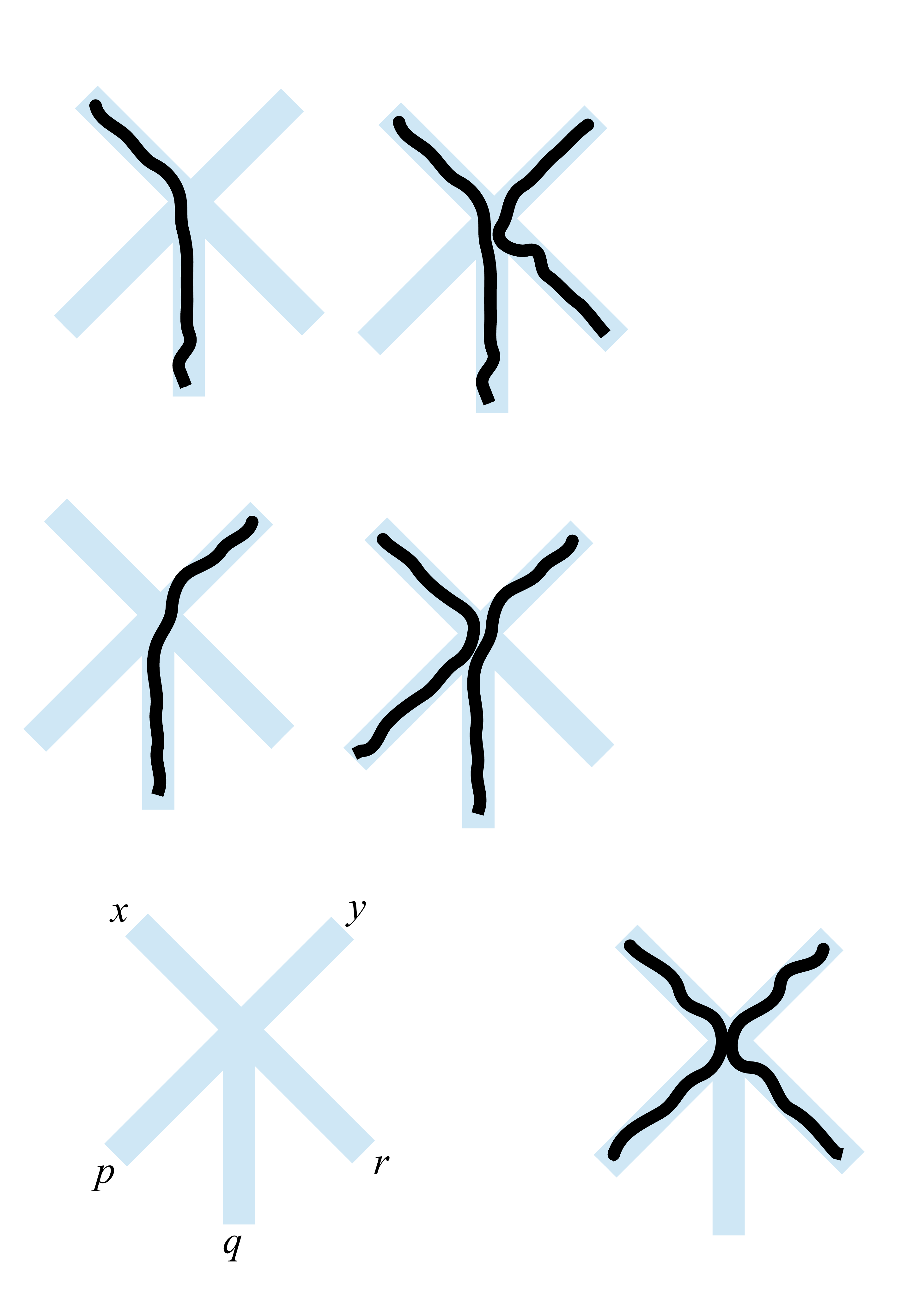}}\\
\subfigure[]{\includegraphics[scale=0.3]{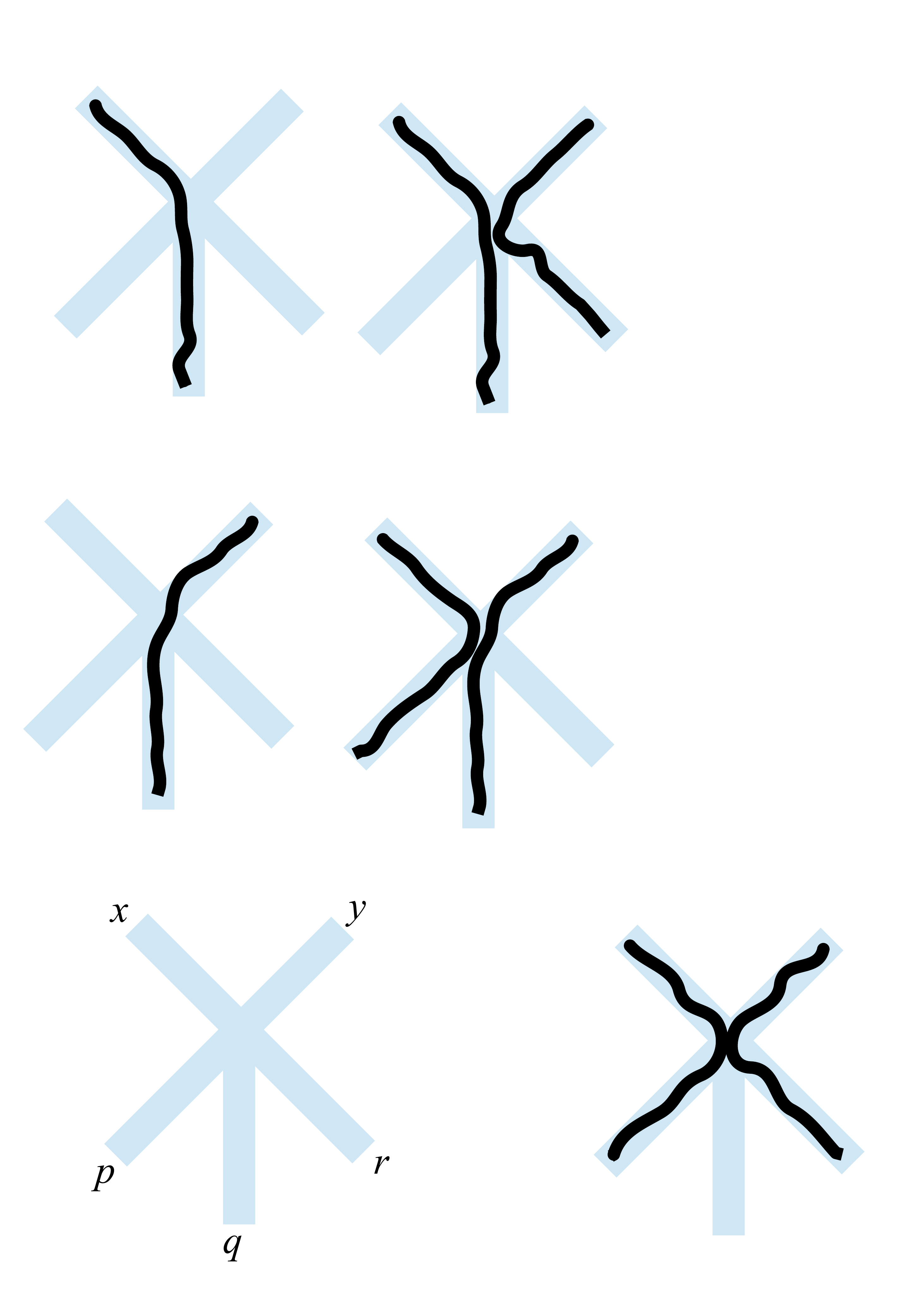}}
\subfigure[]{\includegraphics[scale=0.3]{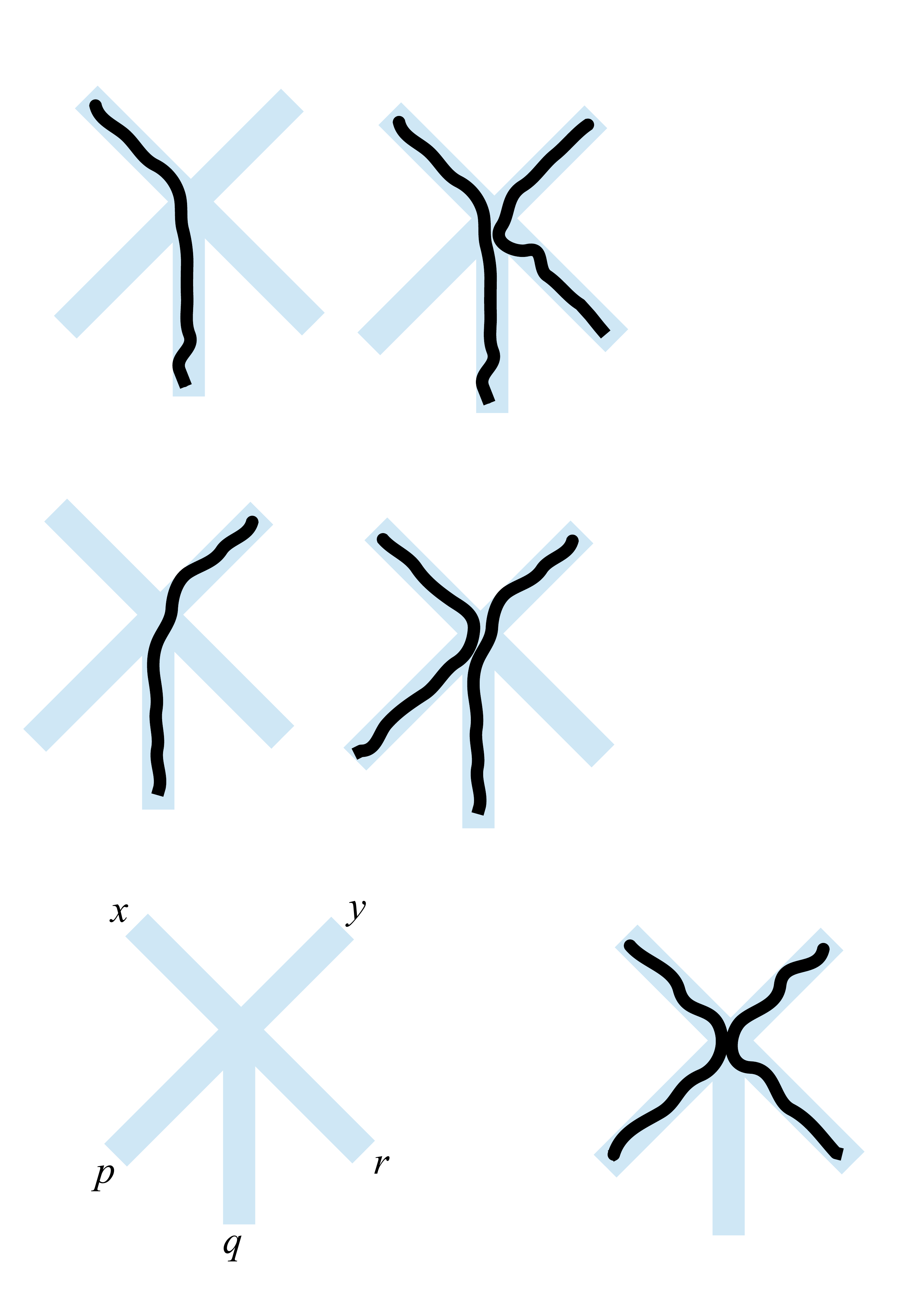}}
\subfigure[]{\includegraphics[scale=0.3]{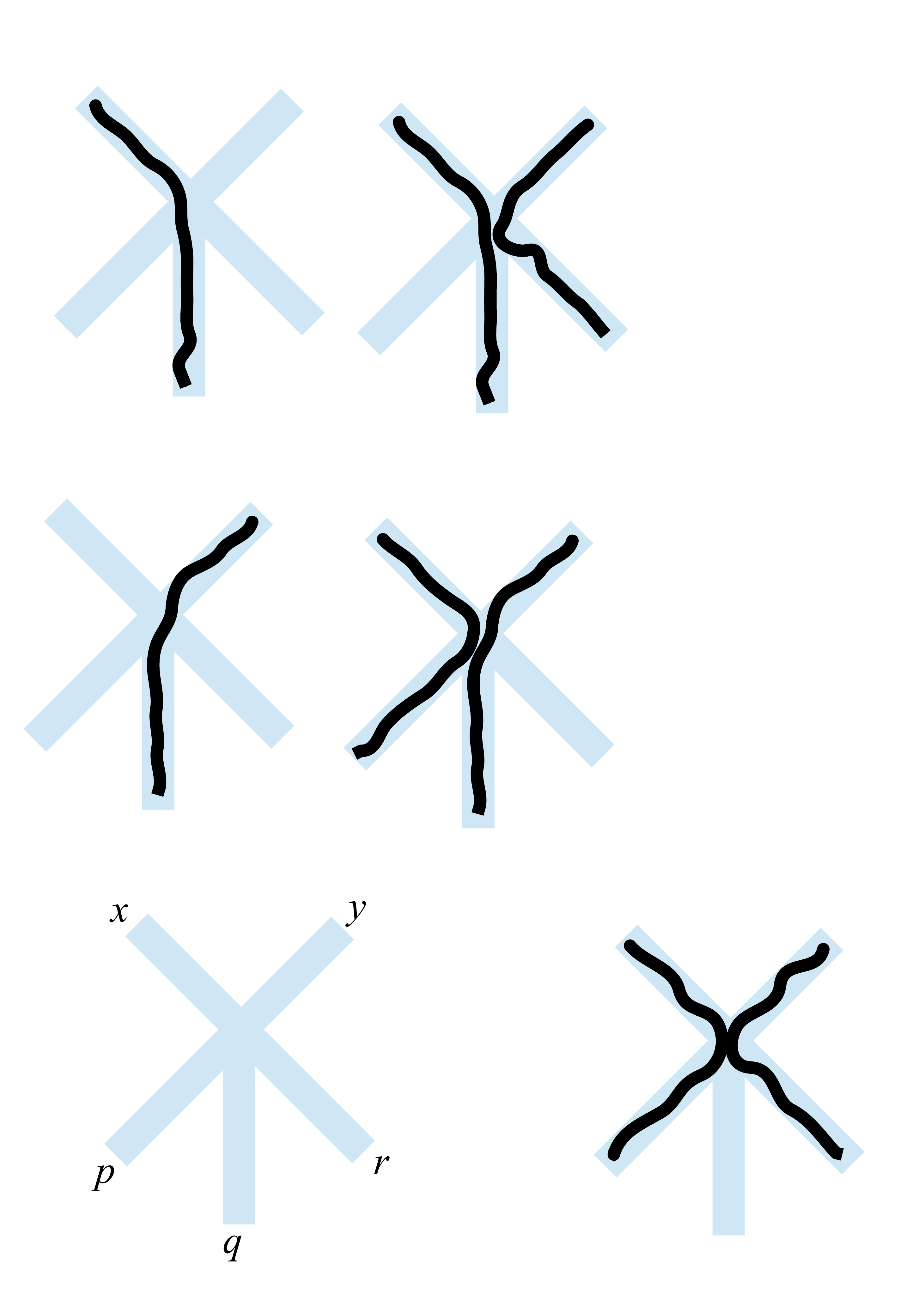}}
\caption{Gravity gate. Plant root logical gate. Only gravitropism is taken into account. Two roots can not fit in one channel.
(a)~Scheme: input channels are $x$ and $y$, output channels are $p$, $q$, $r$. 
(b)~$x=0$, $y=1$. 
(c)~$x=1$, $y=0$. 
(d)~$x=1$, $y=1$, apexes arrive at the junction at the same time.
(e)~$x=1$, $y=1$, apex $x$ arrives at the junction earlier than apex $y$.
(f)~$x=1$, $y=1$, apex $y$ arrives at the junction earlier than apex $x$.
 }
\label{basicgate}
\end{figure}

Consider an interaction gate with two inputs $x$ and $y$ and three outputs $p$, $q$, $r$ (Fig.~\ref{basicgate}a). If channels would be wide enough to accommodate several routes then 
the routes would join each other following along the channel $q$, due to the roots swarming behaviour~\cite{ciszak2012swarming}.  We assume a channel can accommodate only one root.  
Root apexes are guided by gravity. A root entering channel $y$ propagates till the junction, then follows the gravity and moves along channel $q$ (Fig.~\ref{basicgate}b). 
A root entering channel $x$ also propagates along channel $q$ (Fig.~\ref{basicgate}c).  

What happens when two roots enter channels $x$ and $y$ at the same{time}? Assuming roots' apexes reach the junction precisely at the same time they might reflect into lateral channels because two roots at once can not fit in  the vertical channel $q$ (Fig.~\ref{basicgate}c). 
That is an ideal situation. Unlikely this will ever happen because there are no two seeds which produce roots with exactly the same biochemical and physiological parameters. 

In reality one of the roots is faster or stronger. The stronger root  pushes its way into the channel $q$ while contender is  left to deviate into the later channel.  This is illustrated in 
Figs.~\ref{basicgate}ef. If root $x$ wins its way into the channel $q$ then root $y$ grows into the channel $r$. Vice verse, if the root $y$ is quicker to get into channel $q$ 
then root $x$ moves into the channel $p$. Assuming a 
presence of a root in channel $z$ symbolises logical truth: $z=1$, and absence logical false: $z=0$, the gate in Fig.~\ref{basicgate} computes the following Boolean functions: $p=r=xy$ and $q=x\overline{y}+\overline{x}y+xy={x}+y$. However, such a gate is not cascadable 
because --- due to unpredictability of the competition between the apexes for the channel $q$  --- we never know where signal $xy$ appears: either on channel $p$ or on channel $r$.

\begin{figure}[!tbp] 
\centering
\includegraphics[]{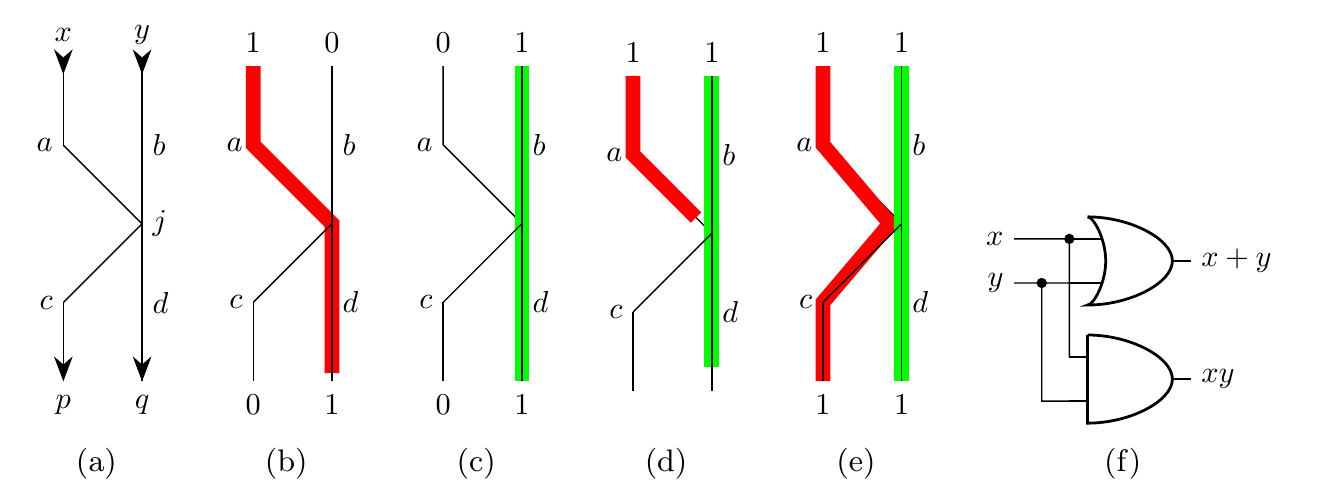}
\caption{Gravity gate with two inputs and two outputs.
    (a)~Scheme of the gate, $p=xy$ and $q=x+y$. 
   {In subfigures (bcde)~ the development of gate is provided}. 
    (b)~Root enters  input channel $x$ only, $x=1$ and $y=0$.
    (c)~Root enters  input channel $y$ only, $x=0$ and $y=1$.
     (d) and (e)~Roots enter both input channels $x$ and $y$, $x=1$ and $y=1$, at different time steps:
    (d)~Time moment $t=b+\sqrt{a}-0.2$, (e)~final state.
    (f)~Equivalent {logic gates design}. 
    Coloured inputs symbolise channels occupied by roots. Red and green colour are only here 
    to show how roots originated at different inputs interact.  
    }
  \label{2I2O_gate}
\end{figure}

To achieve a certainty we should allow one --- specified a priori --- root to reach the junction early. This is how we came up with the gate shown in  Fig.~\ref{2I2O_gate}a.

\begin{table}[!tbp]
\caption{Operations implemented by the gravity gate (Fig.~\ref{2I2O_gate}).}
\centering
\begin{tabular}{c|c||c|c|l}
$x$  & $y$ &  $x+y$ &  $xy$  & Interaction of roots \\ \hline
0  & 0 &   0    & 0     & no roots entered input channels $x$ and $y$ \\
1  & 0 &   1    & 0    & root propagated via channel $x$ to $q$ \\
0  & 1 &   1    & 0    & root propagated via channel $y$ to $q$  \\
1  & 1 &   1    & 1    & root $x$ is blocked by root $y$ and therefore deflected to channel $p$ \\
\end{tabular}
\label{2I2O_gate_table}
\end{table}%

 A channel segment $a$  --- from entry of the input channel $x$ to the junction $j$ --- is longer than segment $b$ --- from the input to $y$ to the junction.  If a root is present only in input $x$, $x=1$, it grows along $a$, reaches the junction and then propagates along $d$  (Fig.~\ref{2I2O_gate}b). 
If a root is present in input $y$, $y=1$, it grows along $b$ and continues along $d$  (Fig.~\ref{2I2O_gate}c). If roots are present in both inputs, $x=1$ and $y=1$, the $y$-root occupies the junction well before the $x$-root reaches the junction (Fig.~\ref{2I2O_gate}d). Thus the $y$-root grows along $d$ while $x$-root reflects into channel $c$ (Fig.~\ref{2I2O_gate}e). The operations are summarised in 
Tab.~\ref{2I2O_gate_table}. The gate realises $xy$ on one output and $x+y$ on another output; an equivalent logic gates design is shown in (Fig.~\ref{2I2O_gate}f). The gate allows for some asynchronicity. It does not matter for how long signal $x$ is delayed, if it enters the circuit at the same time as or any time later than signal $y$ the gate will produce desirable results.

\begin{figure}[!tbp] 
\centering
\includegraphics[]{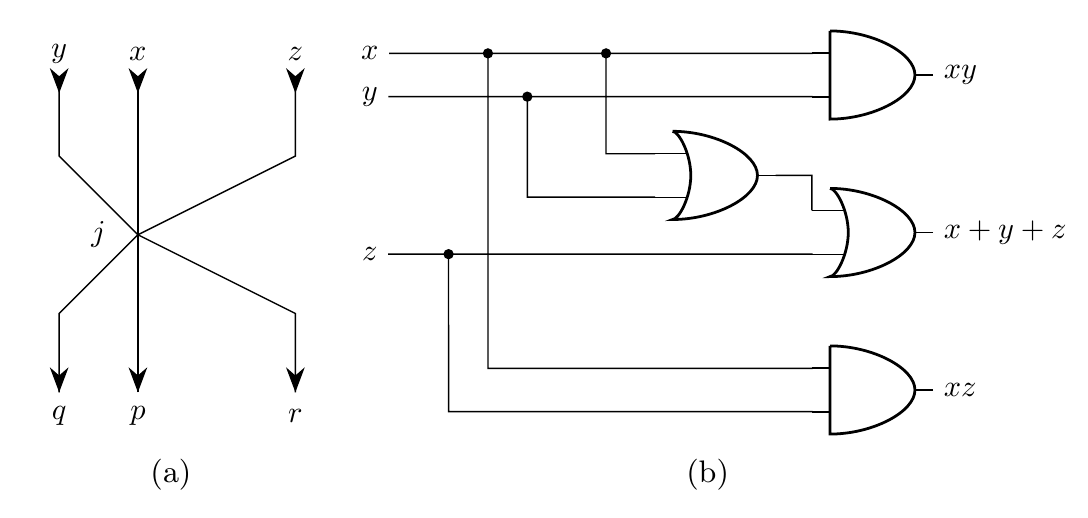}
\caption{(a)~Three-inputs-three-outputs gravity gate: $q=xy$ and $p=x+y+z$ and $r=z(x+y)$ and (b)~its equivalent logical scheme.}
\label{XyZ}
\end{figure}

\begin{table}[!tbp]
\caption{Operations implemented by the three-input-three-output gravity gate (Fig.~\ref{XyZ}).}
\centering
\begin{tabular}{c|c|c||c|c|c|l}
$x$&  $y$ &  $z$ & $x+y+z$ &  $xy$ & $xz$ &    Interaction of roots  \\ \hline
0 &     0 &  0  &    0     &  0   & 0   &     no roots enter input channels   \\
0 &     0 &  1  &    1     &  0  &  0   &   root grows in channel $z$ and exits via channel $p$  \\
0 &     1 &  0  &    1     &  0  &  0   &   root grows in $y$ and exits into $p$    \\
0 &     1 &  1  &    1     &  0  &  0   &   root grows in $y$ and enters $p$, while root in $z$ is reflected into $r$  \\
1 &     0 &  0  &    1     &  0  &  0   &   root grows in $x$ and exits into $p$ \\
1 &     0 &  1  &    1     &  0  &  1   &   root grows in $x$ and exits into $p$ and root in $z$ is reflected into $r$  \\
1 &     1 &  0  &    1     &   1  &  0  &   root grows in $x$ and exits into $p$ and root in $y$ is reflected into $q$\\
1 &     1 &  1   &   1  &  1  &  1   &      root grows in $x$ and exits into $p$, root in $y$ and $z$ are reflected \\
 & & & & & & one after the other, i.e. root in $y$ is reflected into $q$, and root in \\
 & & & & & &  $z$ is reflected into $r$\\
\end{tabular}
\label{XyZ_table}
\end{table}%

The gravity gate $\langle x, y \rangle \rightarrow \langle xy, x+y\rangle$ (Fig.~\ref{2I2O_gate}) can be extended into three-inputs-three outputs gate shown in Fig.~\ref{XyZ}a. The lengths of input channels are selected so that $x$-root reaches the junction earl{ier} than $y$-root, and $y$-root reaches the junction earl{ier} than $z$-root. 
Path along channel $x$ to junction $j$ is shorter than path along channel $y$ to junction $j$. Path along channel $y$ to junction $j$ is shorter than path along channel $z$ to junction $j$.

A root appears in the output channel $p$ if a root grows at least in one of the input channels $x$, $y$ or $z$, respectively. A root appears in the output channel $q$ if roots are initiated in input channels $x$ and $y$. A root appears in output $r$ only if roots grow in channels $x$ and $z$, or in channels $y$ and $z$. 
This gate realises Boolean functions $q=xy$, $p=x+y+z$, and $r=z(x+y)$ (Fig.~\ref{XyZ}b). Operations implemented by the gate are explained in Tab.~\ref{XyZ_table}.

\section{Attraction gates}

Root apexes are attracted to humidity~\cite{baluvska2004root} and a range of chemical compounds~\cite{schlicht2013indole, xu2013improved, graham1991flavonoid, bais2006role, steinkellner2007flavonoids, yokawa2014binary}.  A root propagates towards the domain with highest concentration of attractants. The root minimises energy during its growth: it does not change its velocity vector if environmental conditions stay the same. This is a distant analog of inertia.

\begin{figure}[!tbp] 
\centering
\includegraphics[]{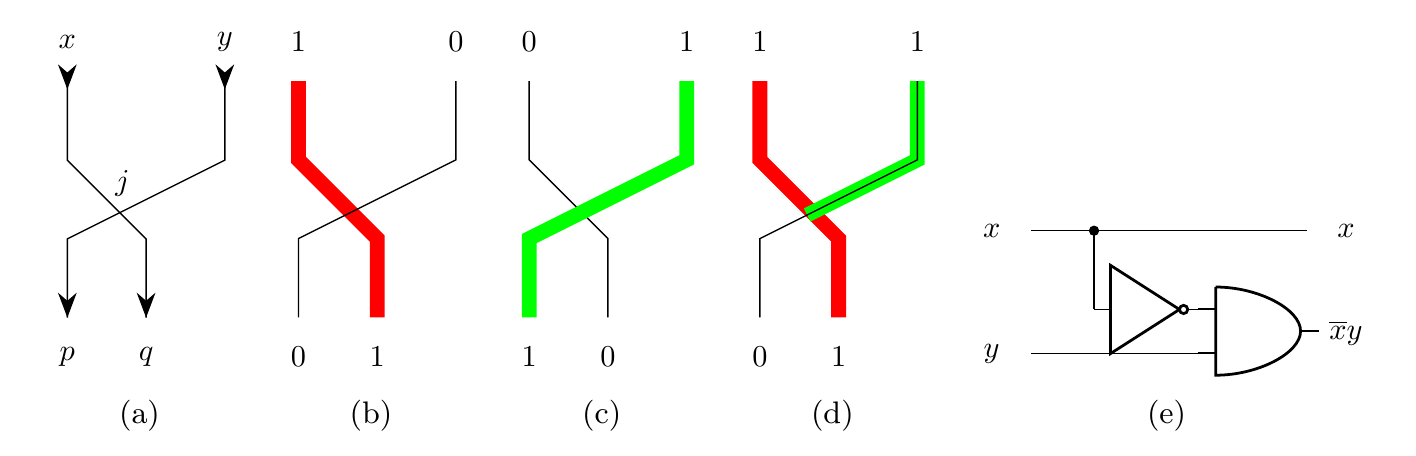}
\caption{(a)~Scheme of humidity gate with two inputs $x$ and $y$ and two outputs $p$ send $q$: $p=\overline{x}y$ and $q=x$.
(b)~$x=1$ and $y=0$. 
(c )~$x=0$ and $y=1$. 
(d)~$x=1$ and $y=1$.
(e)~Equivalent logic scheme.}
\label{humidity}
\end{figure}

Assume attractants are applied
at the exits of channels $p$ and $q$ (Fig.~\ref{humidity}a). When an apex of the root growing along channel $x$ reaches a junction between channels, the apex continues (due to energy minimisation) its growth into the channel $q$ if this channel is not occupied by other root (Fig.~\ref{humidity}b). A root in input channel $y$ grows through the junction into the output channel $p$ (Fig.~\ref{humidity}c).

The gate Fig.~\ref{humidity}a has such a geometry that a path along channel $x$ to  junction $j$ is shorter than a path along channel $y$ to the junction $j$. Therefore, a root growing in channel $x$ propagates through the junction into channel $q$ before root starting in channel $y$ reaches the junction. When both roots are initiated in the input channels the $x$-root appears in the output $q$ but the $y$-root is blocked by the $x$-root from propagating into the channel $p$: no signal appears at the output $p$ (Fig.~\ref{humidity}d). 
This gate realises  functions $p=\overline{x}y$ and $q=x$ (Fig.~\ref{humidity}e).  If $y$ is always 1 the gate produces a signal and its negation at the same time.

\begin{figure}[!tbp] 
\centering
\includegraphics[]{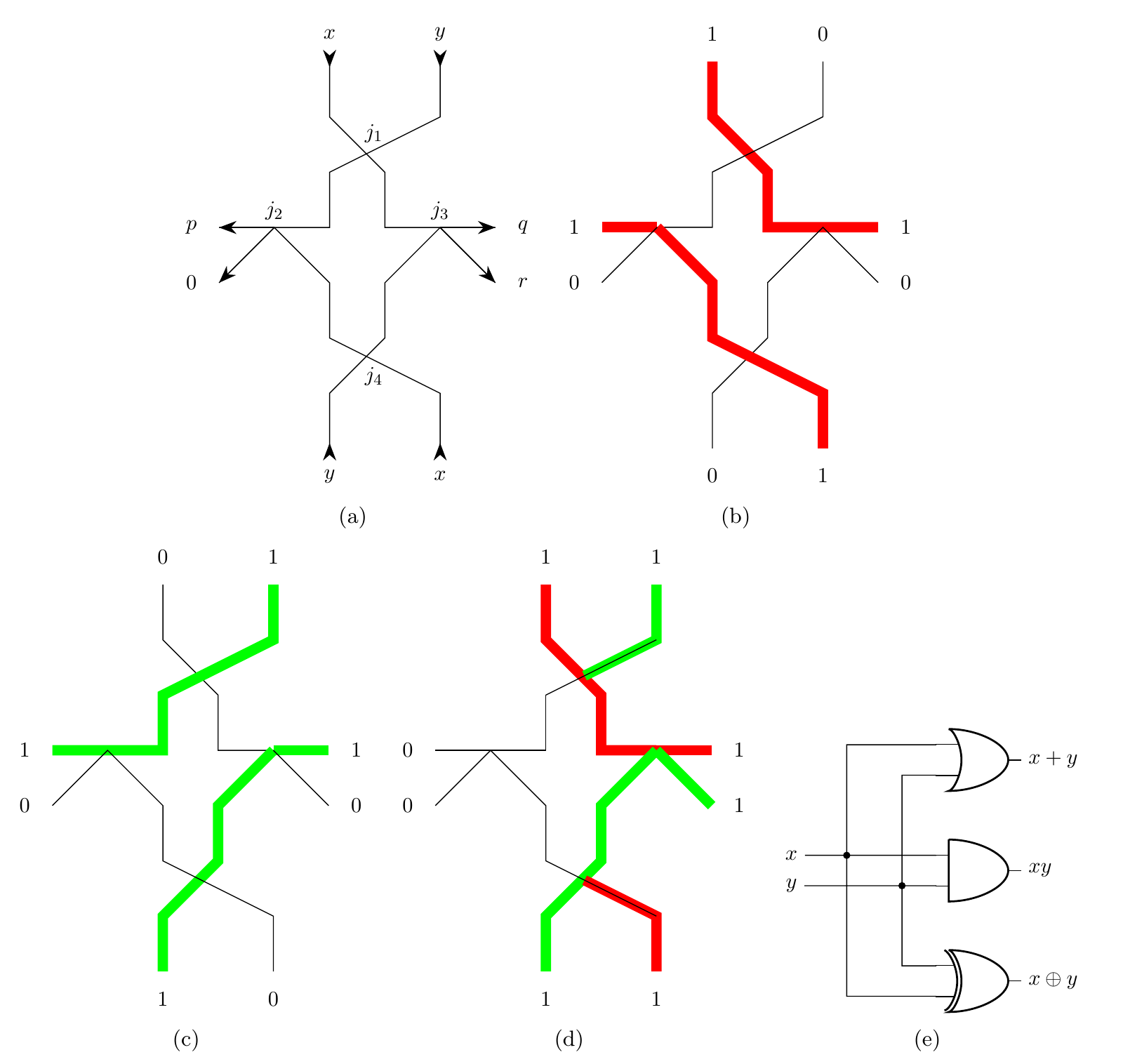}
 \caption{A half-adder made of two humidity gates. 
 (a)~Scheme of the circuit, $p= x \oplus y$, $q=x+y$, $r=xy$.
 (b)~$x=1$ and $y=0$. 
(c )~$x=0$ and $y=1$. 
(d)~$x=1$ and $y$.
(e)~Equivalent logic gates design.}
\label{humidityadder1}
\end{figure} 

Two attraction gates Fig.~\ref{humidity}a  can be cascaded into a circuit to implement a one-bit half-adder, with additional output, 
as shown in Fig.~\ref{humidityadder1}a. We assume the planar gate is lying flat and sources of attractants are provided near 
exits of output channels $p$, $q$ and $r$. The half-adder is realised on inputs $p=x \oplus y$ (sum) and $r=xy$ (carry); 
the circuit has also a  `bonus' output $q=x+y$ (Fig.~\ref{humidityadder1}e). The circuits work 
 as follows:
\begin{itemize}
\item Inputs $x=1$ and $y=0$: Two roots are initiated in channels marked $x$ in Fig.~\ref{humidityadder1}a; 
one root propagates to junction $j_1$ to junction $j_3$ and exits in channel $q$; 
another root propagates to junction $j_4$ to junction $j_2$ and into channel $p$ (Fig.~\ref{humidityadder1}b). 
\item Inputs $x=0$ and $y=1$: Two roots are initiated in channels marked $y$ in Fig.~\ref{humidityadder1}a;
one root propagates to junction $j_1$ then to junction $j_2$ and exits at channel $p$;
another root propagates to junction $j_4$ then to junction $j_3$ then into channel $q$ (Fig.~\ref{humidityadder1}c).
\item Inputs $x=1$ and $y=1$:  Roots are initiated in all four input channels. The root initiated in the northern channel $x$ propagates towards exit $q$. This root 
blocks propagation of the root initiated in the southern channel $y$, therefore the root from the southern channel $x$ exits the circuit via the channel $r$. The root growing 
in the northern channel $x$ blocks propagation of the root initiated in the norther channel $y$, therefore no roots appear in the output $p$ (Fig.~\ref{humidityadder1}d).
\end{itemize}

 \section{Discussion}
 
 The theoretical designs of the plant root gates proposed in the paper wait for their experimental laboratory prototyping. That will be the goal of our further studies in plant-based computing.
 When the prototypes are made they will lay a foundation for a research focused on developing computing architectures from plants, combining bio-electronics, unconventional computing, advanced functional materials, plant biology, robotics.  The paper addressed new trends in computing, especially bio- and nature-inspired by encompassing key aspects of information processing in living plants and adaptation of plants processing structure. The proposed research is tailored to future emerging challenges in living technologies and unconventional computing in highly interdisciplinary settings by developing new kinds of computational approaches in science. Are there any applications apart of making arithmetic-logical units with plant roots? The plant roots logical circuits can be embedded into decision making modules of root-inspired robots  for soil exploration~\cite{mazzolai2010plant, mazzolai2011miniaturized, lucarotti2015revealing, sadeghi2014novel}. The gates can be used as pre-programmed routing devices for automatic manufacturing of plant-based electronic devices, which will incorporate plant wires and memristors~\cite{adamatzky2013physarum, volkov2014memristors, volkov2016biosensors}.

\section*{References}

\bibliographystyle{plain}
\bibliography{plantgatebib}

\begin{thebibliography}{10}

\bibitem{adamatzky2002collision}
Andrew Adamatzky, editor.
\newblock {\em Collision-based computing}.
\newblock Springer, 2002.

\bibitem{adamatzky2009hot}
Andrew Adamatzky.
\newblock Hot ice computer.
\newblock {\em Physics Letters A}, 374(2):264--271, 2009.

\bibitem{adamatzky2010slime}
Andrew Adamatzky.
\newblock Slime mould logical gates: exploring ballistic approach.
\newblock {\em arXiv preprint arXiv:1005.2301}, 2010.

\bibitem{adamatzky2013physarum}
Andrew Adamatzky.
\newblock Physarum wires: self-growing self-repairing smart wires made from
  slime mould.
\newblock {\em Biomedical Engineering Letters}, 3(4):232--241, 2013.

\bibitem{adamatzky2014slime}
Andrew Adamatzky and Theresa Schubert.
\newblock Slime mold microfluidic logical gates.
\newblock {\em Materials Today}, 17(2):86--91, 2014.

\bibitem{bais2006role}
Harsh~P Bais, Tiffany~L Weir, Laura~G Perry, Simon Gilroy, and Jorge~M Vivanco.
\newblock The role of root exudates in rhizosphere interactions with plants and
  other organisms.
\newblock {\em Annu. Rev. Plant Biol.}, 57:233--266, 2006.

\bibitem{baluvska1997root}
F~Balu{\v{s}}ka and KH~Hasenstein.
\newblock Root cytoskeleton: its role in perception of and response to gravity.
\newblock {\em Planta}, 203(1):S69--S78, 1997.

\bibitem{baluvska1996gravitropism}
Franti{\v{s}}ek Balu{\v{s}}ka, Martin Hauskrecht, Peter~W Barlow, and Andreas
  Sievers.
\newblock Gravitropism of the primary root of maize: a complex pattern of
  differential cellular growth in the cortex independent of the microtubular
  cytoskeleton.
\newblock {\em Planta}, 198(2):310--318, 1996.

\bibitem{baluvska2016vision}
Frantisek Balu{\v{s}}ka and Stefano Mancuso.
\newblock Vision in plants via plant-specific ocelli?
\newblock {\em Trends in Plant Science}, 21(9):727--730, 2016.

\bibitem{baluvska2004root}
Franti{\v{s}}ek Balu{\v{s}}ka, Stefano Mancuso, Dieter Volkmann, and Peter
  Barlow.
\newblock Root apices as plant command centres: the unique
  �brain-like�status of the root apex transition zone.
\newblock {\em Biologia (Bratisl.)}, 59(Suppl. 13):1--13, 2004.

\bibitem{baluvska2009root}
Franti{\v{s}}ek Balu{\v{s}}ka, Stefano Mancuso, Dieter Volkmann, and Peter
  Barlow.
\newblock The �root-brain�hypothesis of charles and francis darwin: revival
  after more than 125 years.
\newblock {\em Plant signaling \& behavior}, 4(12):1121--1127, 2009.

\bibitem{baluvska2010root}
Franti{\v{s}}ek Balu{\v{s}}ka, Stefano Mancuso, Dieter Volkmann, and Peter~W
  Barlow.
\newblock Root apex transition zone: a signalling--response nexus in the root.
\newblock {\em Trends in plant science}, 15(7):402--408, 2010.

\bibitem{brenner2006plant}
Eric~D Brenner, Rainer Stahlberg, Stefano Mancuso, Jorge Vivanco,
  Franti{\v{s}}ek Balu{\v{s}}ka, and Elizabeth Van~Volkenburgh.
\newblock Plant neurobiology: an integrated view of plant signaling.
\newblock {\em Trends in plant science}, 11(8):413--419, 2006.

\bibitem{ciszak2012swarming}
Marzena Ciszak, Diego Comparini, Barbara Mazzolai, Frantisek Baluska, F~Tito
  Arecchi, Tam{\'a}s Vicsek, and Stefano Mancuso.
\newblock Swarming behavior in plant roots.
\newblock {\em PLoS One}, 7(1):e29759, 2012.

\bibitem{costello2005experimental}
Benjamin De~Lacy Costello and Andrew Adamatzky.
\newblock Experimental implementation of collision-based gates in
  belousov--zhabotinsky medium.
\newblock {\em Chaos, Solitons \& Fractals}, 25(3):535--544, 2005.

\bibitem{darwin1899geotropism}
Francis Darwin et~al.
\newblock On geotropism and the localization of the sensitive region1 with
  plate xxix.
\newblock {\em Annals of Botany}, 13(4):567--574, 1899.

\bibitem{fredkin2002conservative}
Edward Fredkin and Tommaso Toffoli.
\newblock Conservative logic.
\newblock In Andrew Adamatzky, editor, {\em Collision-Based Computing}, pages
  47--81. Springer, 2002.

\bibitem{gagliano2012towards}
Monica Gagliano, Stefano Mancuso, and Daniel Robert.
\newblock Towards understanding plant bioacoustics.
\newblock {\em Trends in plant science}, 17(6):323--325, 2012.

\bibitem{graham1991flavonoid}
Terrence~L Graham.
\newblock Flavonoid and isoflavonoid distribution in developing soybean
  seedling tissues and in seed and root exudates.
\newblock {\em Plant physiology}, 95(2):594--603, 1991.

\bibitem{gunji2011robust}
Yukio-Pegio Gunji, Yuta Nishiyama, Andrew Adamatzky, Theodore~E Simos, George
  Psihoyios, Ch~Tsitouras, and Zacharias Anastassi.
\newblock Robust soldier crab ball gate.
\newblock {\em Complex systems}, 20(2):93, 2011.

\bibitem{lucarotti2015revealing}
Chiara Lucarotti, Massimo Totaro, Ali Sadeghi, Barbara Mazzolai, and Lucia
  Beccai.
\newblock Revealing bending and force in a soft body through a plant root
  inspired approach.
\newblock {\em Scientific reports}, 5, 2015.

\bibitem{margolus2002universal}
Norman Margolus.
\newblock Universal cellular automata based on the collisions of soft spheres.
\newblock In Andrew Adamatzky, editor, {\em Collision-based computing}, pages
  107--134. Springer, 2002.

\bibitem{martinez2010computation}
Genaro~J Mart{\'\i}nez, Andrew Adamatzky, Kenichi Morita, and Maurice
  Margenstern.
\newblock Computation with competing patterns in life-like automaton.
\newblock In {\em Game of Life Cellular Automata}, pages 547--572. Springer,
  2010.

\bibitem{masi2008electrical}
E~Masi, M~Ciszak, S~Mugnai, E~Azzarello, C~Pandolfi, L~Renna, G~Stefano,
  B~Voigt, D~Volkmann, and S~Mancuso.
\newblock Electrical network activity in plant roots under gravity-changing
  conditions.
\newblock {\em Journal of Gravitational Physiology}, 15(1):167--168, 2008.

\bibitem{mazzolai2008inspiration}
Barbara Mazzolai, Paolo Corradi, Alessio Mondini, Virgilio Mattoli, Cecilia
  Laschi, Stefano Mancuso, Sergio Mugnai, and Paolo Dario.
\newblock Inspiration from plant roots: a robotic root apex for soil
  exploration.
\newblock {\em Proceedings of Biological Approaches for Engineering, University
  of Southampton}, pages 50--3, 2008.

\bibitem{mazzolai2010plant}
Barbara Mazzolai, Cecilia Laschi, Paolo Dario, Sergio Mugnai, and Stefano
  Mancuso.
\newblock The plant as a biomechatronic system.
\newblock {\em Plant signaling \& behavior}, 5(2):90--93, 2010.

\bibitem{mazzolai2011miniaturized}
Barbara Mazzolai, Alessio Mondini, Paolo Corradi, Cecilia Laschi, Virgilio
  Mattoli, Edoardo Sinibaldi, and Paolo Dario.
\newblock A miniaturized mechatronic system inspired by plant roots for soil
  exploration.
\newblock {\em IEEE/ASME Transactions on Mechatronics}, 16(2):201--212, 2011.

\bibitem{mo2015and}
Mei Mo, Ken Yokawa, Yinglang Wan, and Franti{\v{s}}ek Balu{\v{s}}ka.
\newblock How and why do root apices sense light under the soil surface?
\newblock {\em Frontiers in plant science}, 6, 2015.

\bibitem{morgan2016simple}
Alex~JL Morgan, David~A Barrow, Andrew Adamatzky, and Martin~M Hanczyc.
\newblock Simple fluidic digital half-adder.
\newblock {\em arXiv preprint arXiv:1602.01084}, 2016.

\bibitem{pfeffer1894geotropic}
Wilhelm Pfeffer.
\newblock Geotropic sensitiveness of the root-tip.
\newblock {\em Annals of Botany}, 8(31):317--320, 1894.

\bibitem{sadeghi2014novel}
Ali Sadeghi, Alice Tonazzini, Liyana Popova, and Barbara Mazzolai.
\newblock A novel growing device inspired by plant root soil penetration
  behaviors.
\newblock {\em PloS one}, 9(2):e90139, 2014.

\bibitem{schlicht2013indole}
Markus Schlicht, Jutta Ludwig-M{\"u}ller, Christian Burbach, Dieter Volkmann,
  and Frantisek Baluska.
\newblock Indole-3-butyric acid induces lateral root formation via
  peroxisome-derived indole-3-acetic acid and nitric oxide.
\newblock {\em New Phytologist}, 200(2):473--482, 2013.

\bibitem{steinkellner2007flavonoids}
Siegrid Steinkellner, Venasius Lendzemo, Ingrid Langer, Peter Schweiger,
  Thanasan Khaosaad, Jean-Patrick Toussaint, and Horst Vierheilig.
\newblock Flavonoids and strigolactones in root exudates as signals in
  symbiotic and pathogenic plant-fungus interactions.
\newblock {\em Molecules}, 12(7):1290--1306, 2007.

\bibitem{tsuda2004robust}
Soichiro Tsuda, Masashi Aono, and Yukio-Pegio Gunji.
\newblock Robust and emergent physarum logical-computing.
\newblock {\em Biosystems}, 73(1):45--55, 2004.

\bibitem{volkov2016biosensors}
Alexander~G Volkov.
\newblock Biosensors, memristors and actuators in electrical networks of
  plants.
\newblock {\em International Journal of Parallel, Emergent and Distributed
  Systems}, pages 1--12, 2016.

\bibitem{volkov2014memristors}
Alexander~G Volkov, Clayton Tucket, Jada Reedus, Maya~I Volkova, Vladislav~S
  Markin, and Leon Chua.
\newblock Memristors in plants.
\newblock {\em Plant signaling \& behavior}, 9(3):e28152, 2014.

\bibitem{xu2013improved}
Weifeng Xu, Guochang Ding, Ken Yokawa, Franti{\v{s}}ek Balu{\v{s}}ka, Qian-Feng
  Li, Yinggao Liu, Weiming Shi, Jiansheng Liang, and Jianhua Zhang.
\newblock An improved agar-plate method for studying root growth and response
  of arabidopsis thaliana.
\newblock {\em Scientific reports}, 3:1273, 2013.

\bibitem{yokawa2014binary}
Ken Yokawa and Frantisek Baluska.
\newblock Binary decisions in maize root behavior: Y-maze system as tool for
  unconventional computation in plants.
\newblock {\em IJUC}, 10(5-6):381--390, 2014.

\end{thebibliography}

\end{document}